\def\gte{\lower 0.5ex\hbox{${}\buildrel>\over\sim{}$}}
\def\lte{\lower 0.5ex\hbox{${}\buildrel<\over\sim{}$}}

\documentstyle[11pt,aaspp]{article}
\begin{document}

\title{Formation and Radiation Acceleration of Pair Plasmoids\\
Near Galactic Black Holes}
\author{Hui Li and Edison P. Liang}
\affil{Department of Space Physics and Astronomy, Rice University,
Houston, TX 77251}

\begin{abstract}
We study quantitatively the formation and
radiation acceleration of electron-positron pair plasmoids
produced by photon-photon collisions
near Galactic black holes (GBHs).
The terminal ejecta velocity is found to be completely determined
by the total disk luminosity, proton loading factor
and disk size, with no dependence on the initial velocity.
We discuss potential applications to the recently
discovered Galactic superluminal
sources GRS1915, GROJ1655 and possibly other GBHs.
\end{abstract}


\section{INTRODUCTION}

The origin of relativistic jets from Active Galactic Nuclei (AGNs)
remains one of the great challenges in astrophysics.
Earlier Phinney (1982, 1987) concluded that radiation from
the accretion disk is probably not responsible for accelerating
the superluminal AGN jets
with bulk Lorentz factors $\Gamma \sim 10$ (\cite{por87}).
When $\Gamma \gg 1$,
the aberrated photons in the jet's rest frame comes
from the forward direction against the jet motion, thus
limiting the maximum $\Gamma$ achievable to
only a few, much smaller than the observed values.

Recently, observations of
several Galactic black hole (GBH) candidates have revealed
a causal relation between X-ray and radio
flaring of the central object.
In some cases, radio plasmoids were ejected relativistically
away from the central sources following the x-ray flares
(\cite{mi92,mr94,hjr95}).
In two cases, GRS1915 (\cite{mr94})
and GROJ1655 (\cite{hjr95}), the ejecta motions appear to be superluminal,
resembling the superluminal AGN sources.

Motivated by the correlation of the ejection with the
x-ray flux, energetics of the ejecta and the relatively small
$\Gamma \sim 2.5$, we proposed in a recent paper (\cite{ll95})
a scenario for the origin of the episodic ejections
based on the standard inverse-Compton accretion disk
paradigm (\cite{sle76}).
The idea is that episodic quenching of the (external and
internal synchrotron) soft photon source to a level less
than the internal bremsstrahlung flux
superheats the innermost disk to gamma ray temperatures
(\cite{ld88,wl91,pk95}).
However, these
gamma rays cannot all escape, if their source compactness
($\propto$ gamma ray luminosity/source size, \cite{sv84,zz84})
is too high.  Instead most gamma rays are
converted into pairs via photon-photon pair creation
which will be further accelerated axially by the
radiation pressure of the disk x-ray flux and attain
significant terminal Lorentz factors in a bipolar outflow
if the disk luminosity is
sufficiently high.  When the ejecta encounters the
interstellar medium and develops shocks and turbulence it will
convert the bulk kinetic energy into
high energy particles, emitting the observed radio synchrotron radiation.

In this paper, we summarize some key results
for pair plasma formation
and their radiation acceleration by the disk photons for GBHs.

\section{COMPACTNESS AND FORMATION OF PAIR PLASMOIDS}

To study the formation of the pairs produced by
$\gamma\gamma$ and $\gamma$x collisions {\it outside} the disk,
we assume a 2-zone Keplerian disk model depicted in Fig.1.
We define $r_{\gamma}$ as the boundary, interior (exterior) to which
local bremsstrahlung flux is higher (lower) than the soft
photon flux. Gamma rays are emitted by a superheated
ion torus within $r_{\gamma}$ (\cite{rees82,liang90}),
The spectrum from this region is assumed to be a
Comptonized bremsstrahlung + annihilation spectrum
with $T_{\gamma} \sim mc^2$ (Fig.1a, e.g. \cite{zz84,ld88,ku88}).
Exterior to $r_{\gamma}$, we assume that the disk x-ray flux
has the standard inverse-Compton spectral shape
(Fig.1b, \cite{sle76,st80}).
Using standard photon-photon pair production
cross-section we compute the (angle-averaged)
gamma ray pair-production depth
$\tau_{\gamma\gamma}$ as a function of photon energy
due to collisions among
gamma rays and between gamma rays and disk x-rays:
\begin{equation}
\tau_{\gamma\gamma} \simeq r_{\gamma}~
\int d\epsilon^{\prime} n(\epsilon^{\prime})
\sigma_{\gamma\gamma}(\epsilon,\epsilon^{\prime}) f
\end{equation}
\noindent where $\sigma_{\gamma\gamma}(\epsilon,\epsilon^{\prime})$
is the angle-averaged pair production cross-section (\cite{gs67}),
and $\epsilon\equiv h\nu/511$keV.
$n(\epsilon^{\prime})=n_{\gamma}(\epsilon^{\prime})
+n_x(\epsilon^{\prime})$ is the normalized photon spectral
density evaluated at $r_{\gamma}$, and $f$ is a
fudge factor of order unity to
incorporate various complex geometry and angle-dependent effects
(e.g. photon intensity anisotropy and variation
with path length along the line of sight, angular dependence
of $\sigma_{\gamma\gamma}$, etc.).
{\it We calibrate $f$ using sample numerical simulations}
(details to be published elsewhere).
For given geometry and spectral shapes (cf. Fig.1),
$\tau_{\gamma\gamma}$ basically depends only
on the gamma ray source global compactness
$l_{\gamma} = L_{\gamma}/r_{\gamma}/(3.7\times 10^{28}$
erg/cm/s) (\cite{sv84,zz84}).
In Fig.2a we plot $\tau_{\gamma\gamma}$
as functions of photon energy for various $l_{\gamma}$
and $T_{\gamma} \sim mc^2$.  In Fig.2b we plot
the fraction of gamma rays that are absorbed into pairs
as a function of $l_{\gamma}$.
We have assumed that all pairs produced escape without
re-annihilation in situ due to their high kinetic energy
($\sim$ few hundred keV). This is a crude approximation
but we estimate the error to be less than a factor of 2.
The critical compactness at
which $\gte 80\%$ of the gamma rays are
converted into escaping pairs is around
$l_{\gamma} \sim 100$ (with $\gte 2$ uncertainty),
consistent with similar results found by others
(Zdziarski 1995, private communications).
To achieve $l_{\gamma} \geq 100$, Liang \& Li (1995) show that the
overall accretional luminosity $L_0$
must be $\geq 0.1-0.2$ $L_{edd}$.

\section{RADIATION ACCELERATION}

Consider a  plasma of electrons, positrons
and some entrained protons directly above the black hole
 and along the axis of the
disk, interacting with the disk radiation
via only Compton scattering (cf. Fig.1).
Assuming that protons, electrons and positrons are
always co-moving (a cold jet),
$\gamma_{p} \equiv \gamma_{e} = \Gamma$,
and local charge neutrality is
maintained ($n_{e^{-}}=n_{p}+n_{e^{+}}$),
we can obtain the equation of motion for the jet
(at height Z) as:
\begin{equation}
\label{motion}
-{d\Gamma\over dz_*}~=~({m_{e}\over m_{eq}})~\left({r_g\sigma_{T}\over
\beta c}\right)
\int d\epsilon \int d\Omega~
I_{ph}(\epsilon,\Omega) (1-\beta\mu)
[\Gamma^{2}(1-\beta\mu) - 1]
{}~+~{1\over z_*^2}
\end{equation}
\noindent where $z_* = Z/r_g$, $r_g = GM / c^2$, and
$\mu = \cos\theta$,
the angle between the photon momentum and the z-axis, and
$I_{ph}(\epsilon,\Omega)$ is the spectral intensity
seen by the jet
(cf Fig.1, see \cite{od81,ph87} for previous results).
$m_{eq}$ is the equivalent mass of the particles
$= m_{e}+{\zeta_p\over 2-\zeta_p}m_{p}$~with
$\zeta_p=n_{p}/n_{e^{-}}$, the
fraction of proton loading.
For $\zeta_p=0$, $m_{eq}=m_e$~for a pure pair jet and for
$\zeta_p=1$, $m_{eq}=m_{e}+m_{p}$ for an electron-proton jet.
In deriving equation (\ref{motion}), we have:
(1) kept only the parallel (z) component
due to the axial-symmetry of the system
(though there will be a random orthogonal component);
(2) used the angle-dependent differential Compton scattering
cross section (in Thomson limit) and ignored the energy change
(second order in scattered photon energy),
but kept the momentum change
(first order in scattered photon energy) for acceleration;
(3) included the gravity pull on protons and leptons but
neglected the radiation force on protons.

The (isotropic) spectral intensity at the disk surface
can be approximated as:
\begin{equation}
\label{inten}
I_{ph}(\epsilon) = {1.68\times 10^{32}\over \epsilon ~
{\rm exp}\left({511\over 80}\epsilon\right)}
\left({L_0\over L_{edd}}\right)
{1\over M_{10}}~{1\over {\cal G}}~
r_*^{-3} \left(1-\sqrt{6\over r_*}\right)
\left({1\over {\rm cm^2\cdot s \cdot sr}}\right)
\end{equation}

\noindent where $r_* = r/r_g$, $M_{10} = M/10M_{\odot}$ and
${\cal G} = 18\left({1\over r^*_{min}} - {1\over r^*_{max}}\right)
+ 2\left[ \left({6\over r^*_{max}}\right)^{3/2} -
\left({6\over r^*_{min}}\right)^{3/2} \right]$.
Here, $r^*_{min}$ and $r^*_{max}$ define the size of the disk
from which the total luminosity is $L_0$,
and we assume $r^*_{min} = r_{\gamma}/r_g$.
We have used the observed spectrum of GRS1915 (\cite{har94}),
but results are insensitive to detailed spectral shape since it
will be integrated away when putting into equation (\ref{motion})
and only the total flux from each radius matters since we are
working in the Thomson limit (as emphasized by the referee).
The $\gamma$-ray component from $r < r_{\gamma}$ (Fig.1a) produces
negligible additional effects on the axial acceleration due to
the strong $\gamma-\gamma$ self-absorption at high $l_{\gamma}$.
Substituting equation (\ref{inten}) into equation (\ref{motion})
the integral over $\Omega$
becomes an integral over $r_*$ (cf. Fig.1).

Equation (\ref{motion}) can be evaluated numerically
for the terminal $\Gamma_{\infty}$.
A key term in this equation is
$\Gamma^{2}(1-\beta\mu)~-~1$, which signifies the accelerating
(or decelerating) force from a ``ring'' of radiation at radius
$r_*$, depending on whether it is negative (or positive) for particles
at the height $z_*$, where $\mu = z_*/\sqrt{z_*^2+r_*^2}$.
One can find a disk radius $r_{sep}$,
interior to which disk radiation accelerates
the ejecta, and exterior to which radiation decelerates the ejecta.
Whether an ejecta can get net acceleration depends on the sum of
these two contributions.

The dependence of $\Gamma_{\infty}$
on the initial conditions ($\Gamma_{initial}~{\rm and}~z^*_{min}$)
and the geometry ($r^*_{min}~{\rm and}~r^*_{max}$)
can be obtained by solving equation (\ref{motion}).
We will focus on cases with $\zeta_p = 0$ first.
The results are shown in Figure 3 and summarized here.

(A). For fixed disk geometry and total disk
luminosity, {\em regardless of their initial velocities
$\Gamma_{initial}$}, leptons will be accelerated to a fixed
terminal $\Gamma_{\infty}$, even though particles with
high $\Gamma_{initial}$ get decelerated first.
The radiation field acts like a
``thermostat'' that regulates the particle's final velocity along the
z-axis. In Fig.3a, we show the ejecta's $\Gamma$
as a function of height $z_*$ with different initial velocities,
whereas fixing $r^*_{min} = 12,~r^*_{max} = 10^4,~z^*_{min} = 10~
{}~{\rm and}~L_0/L_{edd} = 0.2$.

(B). Fig.3b, 3c, and 3d show the ejecta's $\Gamma$ as a function of
height $z_*$ with different $r^*_{min}$, different
$r^*_{max}$ and different initial $z^*_{min}$, respectively.
Again, $L_0/L_{edd} = 0.2$.
It is evident that, $\Gamma_{\infty}$ depends somewhat sensitively
on $r^*_{min}$ since it is mostly the inner disk radiation that
accelerates the ejecta; on the other hand, the dependence on
$r^*_{max}$ is weak as long as it is $\gg 100$ due to the radiation
intensity drops down drastically as $r^*$ increases.
Note from equation (\ref{motion}) that
the rate of acceleration is proportional to
$d\Omega I_{ph}(\epsilon,\Omega)$, which is decreasing as $1/z_*^2$.
So efficient acceleration only occurs
$within~a~height~of~100-1000 r_g$ of the black hole if
the conditions are favorable.
$\Gamma_{\infty}$ is not a sensitive function of
$z^*_{min}$ as long as $z^*_{min} \leq 50$.
Ejecta starting at very high $z_*$ is not of interest here.

Our main results are summarized in Figure 4, where we show the
terminal $\Gamma_{\infty}$ as a function of $L_0/L_{edd}$ for
different proton loading $\zeta_p = n_p/n_{e^-}$ with
$r^*_{min} = 12$, $r^*_{max} = 10^4$ and $z^*_{min} = 10$.
It is evident that for $\zeta_p > 10^{-3}$, gravitational attraction
on the protons hinders the acceleration.
This has important implications for the composition of the ejecta.
If the jets are composed of pure pairs as suggested in previous sections,
then radiation acceleration can indeed account for the observed
$\Gamma$ in both GRS1915 and GROJ1655 (\cite{mr94,hjr95}).
The observed $\Gamma \sim 2.55$ of GRS1915 implies
$L_0/L_{edd} \sim 0.2$. This constrains the mass of black hole
in GRS1915 to $\sim 12 M_{\odot}$ when using the
observed luminosity (\cite{har94}).

The jet could be suffocated if there is an additional
isotropic radiation field $I_{iso}$ and it reaches
$\sim 10\%$ of $I_{disk}$.  $I_{iso}$ could
be caused by the companion wind or circumstellar medium.

\section{DISCUSSIONS}

The pairs are born with average kinetic energy of several
hundred keV. However, we do not expect the ``Compton rocket''
effect (\cite{od81}) to be important to affect the
ejecta's terminal velocity because the disk radiation field will
regulate the particle's final velocity, no matter what the
initial velocity is, as shown in Fig.3a.

We predict that the terminal $\Gamma_{\infty}$ should
increase as $L_0/L_{edd}$ increases. This can be tested using
the ejections from different episodes with the corresponding
radiation luminosities.

To summarize, in the present approach,
we have adopted the {\it thermal} accretion
disk model for the hard x-ray emissions
(as suggested by the spectra of GRS1915, \cite{har94}),
and use that as the basis for the pair production and
bulk acceleration.  Hence the energetics of the jet is constrained
by the pre-ejection accretional X-ray power output (\cite{ll95}).
We find that, for a jet composed mostly of leptons, disk radiation
can accelerate it to a terminal Lorentz $\Gamma_{\infty}$ of a few,
depending on the total disk luminosity and geometry.
However, {\it non-thermal}
processes may be needed to explain the power-law hard X-ray
spectra of some x-ray novae
(including GROJ1655, \cite{har94,kro95}).
Yet both classes of GBHs show relativistic outflows.
Can GBH jets be also generated
by say, purely electrodynamical processes unrelated to the
conventional thermal hard x-ray disk models?
We will explore alternative models in future publications.

\acknowledgments

We thank an anonymous referee for constructive criticisms.
This work is supported by NASA grant NAG 5-1547 and NRL contract No.
N00014-94-P-2020.

\begin{figure}

\caption{Schematic depiction of the accretion disk along with
emissions from the inner region (spectrum (a)) and the outer disk
(spectrum (b)). At high compactness, spectrum (a) is not visible
at infinity since most of it is converted into pairs within
the dashed sphere. These pairs are then pushed out by the disk
radiation with Lorentz factor $\Gamma$ along the axis of the disk. }

\caption{(a) Pair-production optical depth $\tau_{\gamma\gamma}$
as a function of photon energy.
Curves a, b, and c are for compactness
$l_{\gamma} = 12,~100$, and $392$, respectively;
(b) The fraction of gamma rays that are absorbed
$f_{abs}$ as a function of $l_{\gamma}$.
Note that $f_{abs} \sim 80\%$ for $l_{\gamma} = 100$.}

\caption{Ejecta $\Gamma$ as a function of height $z_*$ as solutions
to equation (3) with $L_0/L_{edd}=0.2~{\rm and}~ \zeta_p=0.0$
(pure pair jet).
Parametric dependence of $\Gamma$ on initial velocity $\Gamma_{initial}$,
$r^*_{min}$, $r^*_{max}$ and $z^*_{min}$ are shown in plot (a), (b), (c)
and (d), respectively. In each plot, other than the parameter that
is varying, the rest are chosen from $\Gamma_{initial} = 1$,
$r^*_{min} = 12$, $r^*_{max} = 10^4$ and $z^*_{min} = 10$. }

\caption{The ejecta's terminal $\Gamma_{\infty}$ as a function
of $L_0/L_{edd}$ for different proton loading $\zeta_p = n_p/n_{e^-}$.
Curves $a,b,c,d,e$ correspond to  $\zeta_p=0.0~({\rm pure~pair}),~10^{-3},
10^{-2}, 10^{-1}, 1~({\rm e-p~jet})$.
The insert corresponds to the vertical cuts for
$L_0/L_{edd} = 0.25, 2.5~{\rm and}~25$, respectively.
The observed $\Gamma$ for GRS1915 is $\sim 2.55$.}

\end{figure}


\begin{thebibliography}{}
\bibitem[Gould \& Schreder 1967]{gs67} \reference Gould, R. J. \&
Schreder, G. P. 1967, Phys. Rev., 155, 1404
\bibitem[Harmon et al. 1994]{har94} \reference Harmon, B. A. et al.
in The Second Compton Symposium, ed. C. Fichtel, N. Gehrels \& J. Norris
(New York: AIP Conf. Proc. No. 304), 210
\bibitem[Hjellming \& Rupen 1995]{hjr95} \reference Hjellming,
R. M. \& Rupen, M. P. 1995, Nature, 375, 464
\bibitem[Kroeger et al. 1995]{kro95} \reference Kroeger, R. A. et al.
1995, Nature, {\it submitted}
\bibitem[Kusunose \& Takahara 1988]{ku88} \reference Kusunose, M. \&
Takahara, F. 1988, PASJ, 40, 435
\bibitem[Liang 1990]{liang90} \reference Liang, E. P. 1990, A\&A, 227, 447
\bibitem[Liang \& Dermer 1988]{ld88} \reference Liang, E. P. \& Dermer,
C. D. 1988, \apj, 325, L39
\bibitem[Liang \& Li 1995]{ll95} \reference Liang, E. P. \& Li, H. 1995,
A\&A, 298, L45
\bibitem[Mirabel et al. 1992]{mi92} \reference Mirabel, I. F. et al. 1992,
Nature, 358, 215
\bibitem[Mirabel \& Rodriguez 1994]{mr94} \reference Mirabel, I. F.
\& Rodriguez, L. F. 1994, Nature, 371, 46
\bibitem[O'Dell 1981]{od81} \reference O'Dell, S. L. 1981, \apjlett, 243, L147
\bibitem[Phinney 1982]{ph82} \reference Phinney, E. S. 1982, \mnras, 198, 1109
\bibitem[Phinney 1987]{ph87} \reference Phinney, E. S. 1987, in Superluminal
Radio Sources, ed. J. A. Zensus \& T. J. Pearson (Cambridge: Cambridge
Univ. Press), 301
\bibitem[Pietrini \& Krolik 1995]{pk95} \reference Pietrini, P. \& Krolik,
J. H. 1995, preprint
\bibitem[Porcas 1987]{por87} \reference Porcas, R. W. 1987, in Superluminal
Radio Sources, ed. J. A. Zensus \& T. J. Pearson (Cambridge: Cambridge
Univ. Press), 12
\bibitem[Rees et al. 1982]{rees82} \reference Rees, M. et al. 1982, Nature,
295, 17
\bibitem[Shapiro et al. 1976]{sle76} \reference Shapiro, S. L.,
Lightman, A. P. \& Eardley, D. M. 1976, \apj, 204, 187
\bibitem[Sunyaev \& Titarchuk 1980]{st80} \reference Sunyaev, R. \& Titarchuk,
L. 1980, A\&A, 86, 121
\bibitem[Svensson 1984]{sv84} \reference Svensson, R. 1984, \mnras, 209, 175
\bibitem[Wandel \& Liang 1991]{wl91} \reference Wandel, A. \& Liang, E. P.
1991, \apj, 380, 84
\bibitem[Zdziarski 1984]{zz84} \reference Zdziarski, A. A. 1984, \apj, 283,
842

\end{thebibliography}
\end{document}